\documentclass[twocolumn,aps,showpacs,floatfix,superscriptaddress]{revtex4}
\usepackage{amsmath,amssymb,eucal,graphicx}

\def\Ro#1{${\bf R}^{(2)}_{#1}$}
\def\Re#1{${\bf R}^{(3)}_{#1}$}

\begin{document}

\title{Sequence Nets}

\author{Jie Sun}
\email{sunj@clarkson.edu}
\author{Takashi Nishikawa}
\email{tnishika@clarkson.edu}
\affiliation{Department of Mathematics \& Computer Science,  Clarkson University Potsdam, NY 13699-5815}
\author{Daniel ben-Avraham}
\email{qd00@clarkson.edu}
\affiliation{Department of Physics, Clarkson University, Potsdam, NY 13699-5820}

\begin{abstract}
  We study a new class of networks, generated by sequences of letters taken from a finite alphabet consisting
  of $m$ letters (corresponding to $m$ types of nodes) and a fixed set of connectivity rules.  Recently, it was  shown  
  how a binary alphabet might generate threshold nets in a similar fashion [Hagberg et al., Phys.~Rev.~E 74, 056116 
  (2006)]. Just like threshold nets,  
  sequence 
  nets in general possess a modular structure reminiscent of everyday life nets, and are easy to handle analytically (i.e., calculate degree distribution, shortest paths, betweenness centrality, etc.).
  Exploiting symmetry, we make a full classification of two- and three-letter sequence nets, discovering two new classes of two-letter
  sequence nets.  The new sequence nets retain many of the desirable analytical properties of
  threshold nets while yielding richer possibilities for the modeling of everyday life complex networks more faithfully.
   \end{abstract}
\maketitle

\section{INTRODUCTION}

Threshold nets are obtained by assigning a weight $x$, from a distribution $\rho(x)$, to each of $N$ nodes
and connecting any two nodes $i$ and $j$ whose combined weights exceed a certain threshold, $\theta$: 
$x_i+x_j>\theta$~\cite{cal,bog,mas,kon}.  Threshold nets can be produced of (almost) arbitrary degree distributions, including scale-free, by judiciously choosing the weight distribution $\rho(x)$ and the threshold $\theta$, and they encompass an astonishingly wide variety of important architectures: from the star graph (a simple ``cartoon" model of scale-free graphs --- consisting of a single hub) with its low density of links, $2/N$, to the complete graph.  Studied extensively in the graph-theoretical literature~\cite{gol80,mah,ham,mer}, they have recently come to the attention of statistical and non-linear physicists due to the beautiful work of Hagberg, Swart, and Schult~\cite{hag}.

\begin{figure}[ht]
\includegraphics*[width=0.35\textwidth]{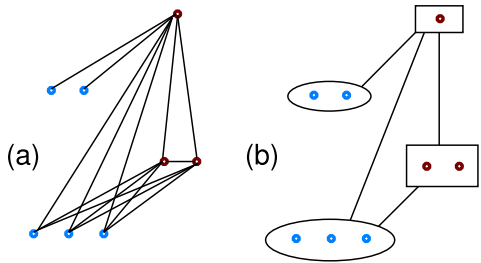}
\caption{Threshold network: (a)~The threshold graph resulting from
the sequence $(A,A,A,B,B,A,A,B)$, and (b)~its box representation, highlighting modularity.  Nodes are added one at a time from bottom to top, $A$'s on the left and $B$'s on the right.}
\label{graph_box}
\end{figure}

Hagberg {\it et al}., exploit the fact that threshold graphs may be more elegantly encoded by a two-letter sequence, corresponding to two types of nodes, $A$ and $B$~\cite{rem1}.
As new nodes are introduced, according to a prescribed sequence, nodes of type $A$ connect to none of the existing nodes, while nodes of type $B$ connect to all of the nodes, of either type: $B\to A$ and $B\to B$.
In Fig.~\ref{graph_box}(a) we show an example of the threshold graph obtained from the sequence $(A,A,A,B,B,A,A,B)$.
Note the {\it modular\/} structure of threshold graphs: a subsequence of $n$ consecutive $B$'s gives rise to
a $K_n$-clique, while nodes in a subsequence of $A$'s connect to $B$ nodes thereafter, but not among one
another.  We highlight this modularity with a diagram of boxes (similar to~\cite{hag}):
oval boxes enclose nodes of type $A$, that are not connected among themselves, while rectangular boxes
enclose $K$-cliques of $B$-nodes~\cite{boxy}.  A link between two boxes means that all of the nodes in one box are connected to all of the nodes in the other, Fig.~\ref{graph_box}(b).

Given the sequence of a threshold net, there exist fast algorithms to compute important structural benchmarks, besides its modularity, such as degree distribution,  triangles, betweenness centrality, and the spectrum and eigenvectors of the graph Laplacian~\cite{hag}.  The latter are a crucial determinant of dynamics and synchronization and have applications to graph partitioning and mesh processing~\cite{bar,nis,hon,hwa,mot,got}.  Perhaps more importantly, it becomes thus possible to {\it design\/} threshold nets with a particular degree distribution, spectrum of eigenvalues, etc.,~\cite{hag}.  

Despite their malleability, threshold nets
are limited in some obvious ways, for example their diameter is 1 or 2, regardless of the number of 
nodes $N$.  Our idea consists of studying the broader class of nets that can be constructed from a sequence
(formed from two or more letters) by deterministic rules of connectivity on their own right.  It is truly this property that gives the nets all their desired attributes: modularity (as in everyday life complex nets), easily computable structural measures --- including the possibility of design --- and a high degree of compressibility.  Roughly speaking, each additional letter to the alphabet allows for an increase of one link in the nets' diameter, so that
the three-letter nets possess diameter 3 or 4 (some of the new types of two-letter nets have diameter 3).  This modest increase is very significant, however, in view of the fact that the diameter of many everyday life complex nets is not much larger than that~\cite{alb}.  Sequence nets gain us
much latitude in the types of nets that can be described in this elegant fashion, while retaining much of the analytical appeal of threshold nets.  Another unusual property of sequence nets is that any
ensemble of sequence nets admits a natural ordering; simply list them alphabetically according to their sequences.
One may use this ordering for exploring eigenvalues and other structural properties of sequence nets. 

In this paper, we make a first stab at the general class of {\it sequence nets\/}.  In Section~\ref{two-letter}
we explore systematically all of the possible rules for creating connected sequence nets from a two-letter alphabet.
Applying symmetry arguments, we find that threshold nets are only one of three equivalence classes, characterized
by the highest level of symmetry.  We then discuss the remaining two classes, showing that also then there is a high
degree of modularity and that various structural properties can be computed easily.  Curiously, the new classes of
two-letter sequence nets can be related to a generalized form of threshold nets, where the {difference} 
$|x_i-x_j|$, rather than
the sum of the weights, is the one compared to the threshold $\theta$.

In Section~\ref{three-letter} we derive all possible forms of connected three-sequence nets.  Symmetry arguments
lead us to the discovery of 30 distinct equivalence classes.  Among these classes, we identify a natural extension of threshold nets to three-letter sequence nets.  Despite the enlarged alphabet, 3-letter
sequence nets do retain many of the desirable properties of threshold and 2-letter sequence nets. 
We also show that at least some of the 3-letter sequence nets can be mapped into 
threshold nets with {\it two\/} thresholds, instead of one.  We conclude with  a summary and discussion of open problems in Section~\ref{conclude}.

\section{2-Letter Sequence Nets}
\label{two-letter}
\subsection{Classification}
%: Classification of 2-letter sequence nets
Consider graphs that can be constructed from sequences $(S_1,S_2,\dots,S_N)$ of the two letters $A$ and $B$.
We can represent any possible rule by a $2\times2$ matrix {\bf R} whose elements
indicate whether nodes of type $i$ connect to nodes of type $j$:
$R_{ij}=1$ if the nodes connect, and 0 otherwise  ($i=1,2$ stands for $A,B$, respectively). Fig.~\ref{graph_box} gives an example of the graph obtained from the sequence $(A,A,A,B,B,A,A,B)$, applying the {\it threshold\/} rule ${0\,0\choose1\,1}$.
Since each element can be $0$ or $1$ independently of the others, there  are $2^4=16$ possible rules.  We shall disregard, however, the four rules that fail to connect between $A$ and $B$,
\begin{equation}
\begin{split}
{\bf R}_0=\left({0\,0\atop0\,0}\right)
,\qquad
{\bf R}_1=\left({1\,0\atop0\,0}\right),  \\
{\bf R}_2=\left({0\,0\atop0\,1}\right)
,\qquad
{\bf R}_3=\left({1\,0\atop0\,1}\right),
\end{split}
\end{equation}
for they yield simple {\it disjoint\/} graphs of the two types of nodes: ${\bf R}_0$ yields isolated nodes only, ${\bf R}_3$ yields one complete graph of type $A$ and one of type $B$, ${\bf R}_1$ yields a complete graph of type $A$ and isolated nodes of type $B$, etc.

\begin{figure}[ht]
\includegraphics*[width=0.35\textwidth]{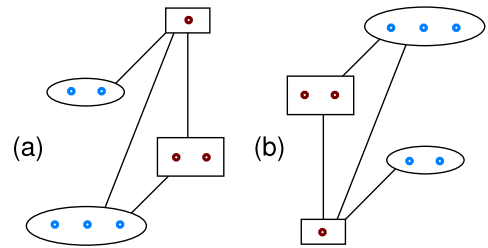}
\caption{Combined time reversal and permutation symmetry: The graphs resulting from
${\bf R}_4$ applied to the sequence $(A,A,A,B,B,A,A,B)$~(a), and from 
${\bf R}_6$
applied to the reverse-inverted sequence $(A,B,B,A,A,B,B,B)$~(b), are identical.}
\label{time_reversal}
\end{figure}

The list of remaining rules can be shortened further by considering two kinds of symmetries: (a)~permutation, and (b)~time reversal.  {\it Permutation\/} is the symmetry obtained by permuting between the two types of nodes, $A\leftrightarrow B$.  Thus, a permuted rule  ($R_{11}\leftrightarrow R_{22}$ and $R_{12}\leftrightarrow R_{21}$) acting on a permuted sequence (${\bar S}_1,{\bar S}_2\dots,{\bar S}_N$) yields back the original graph~\cite{rem2}. {\it Time reversal\/} is  the symmetry obtained by reversing the arrows (``time") in the
connectivity rules, or taking the transpose of ${\bf R}$.  The transposed rule acting on the reversed sequence $(S_N,S_{N-1},\dots,S_1)$  yields back the original graph.  The two symmetry operations are their own inverse and they form a symmetry group.  In particular, one may combine the two symmetries: a rule with $R_{11}\leftrightarrow R_{22}$ applied on a reversed sequence with inverted types
$({\bar S}_N,{\bar S}_{N-1},\dots,{\bar S}_1)$ yields back the original graph, see Fig.~\ref{time_reversal}.   

All of the four rules
\begin{equation}
\begin{split}
{\bf R}_4=\left({0\,0\atop1\,1}\right)
,\qquad
{\bf R}_5=\left({1\,1\atop0\,0}\right),  \\
{\bf R}_6=\left({1\,0\atop1\,0}\right)
,\qquad
{\bf R}_7=\left({0\,1\atop0\,1}\right),
\end{split}
\end{equation}
are equivalent and generate threshold graphs.  ${\bf R}_4$ is the rule for threshold graphs exploited by Hagberg et al.,~\cite{hag}, and ${\bf R}_5$ is equivalent to it by permutation.  ${\bf R}_6$ is obtained from
${\bf R}_4$ by time reversal and permutation (Fig.~\ref{time_reversal}),  and ${\bf R}_7$ is obtained from ${\bf R}_4$ by time reversal.
         
\begin{figure}[ht]
\includegraphics*[width=0.47\textwidth]{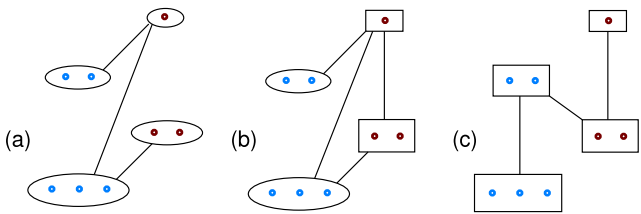}
\caption{Distinct  types of connected non-trivial two-letter sequential graphs:  All three graphs are generated from the same sequence, $(A,A,A,B,B,A,A,B)$, applying
rules ${\bf R}_8$~(a),   ${\bf R}_4$~(b), and   ${\bf R}_{13}$~(c). Note the figure-background symmetry of (a) and (c): the graphs are the inverse, or complement of one another (see text).
The inverse of the threshold graph (b) is also a (two-component) threshold graph, obtained from the same sequence and applying the rule   ${\bf R}_5$  (${\bf R}_4$'s complement).}
\label{ABgraphs}
\end{figure}

The two rules
\begin{equation}
{\bf R}_8=\left({0\,0\atop1\,0}\right)
,\qquad
{\bf R}_9=\left({0\,1\atop0\,0}\right),
\end{equation}
are equivalent, by either permutation or time reversal, and generate non-trivial bipartite graphs that are different from threshold nets (Fig.~\ref{ABgraphs}).  

The rule ${\bf R}_{10}={0\,1\choose1\,0}$ generates complete bipartite graphs.  However, the complete bipartite graph $K_{p,q}$ can also be produced by applying ${\bf R}_8$ to the sequence $(A,A,\dots A,B,B,\dots B)$ of $p$ $A$'s followed by $q$ $B$'s, so the rule ${\bf R}_{10}$ is a ``degenerate'' form of ${\bf R}_8$.  One could see 
that this is the case  at the
outset, because of the symmetrical relations $A\to B$, $B\to A$: these render the ordering of the $A$'s and $B$'s in the graph's sequence irrelevant.  By the same principle, ${\bf R}_{11}={0\,1\choose1\,1}$ and
${\bf R}_{12}={1\,1\choose1\,0}$ are degenerate forms of ${\bf R}_4$ and ${\bf R}_5$, respectively.  They yield threshold graphs with segregated sequences of $A$'s and $B$'s.

The two rules       
\begin{equation}
{\bf R}_{13}=\left({1\,1\atop0\,1}\right)
,\qquad
{\bf R}_{14}=\left({1\,0\atop1\,1}\right),
\end{equation}  
are equivalent, by either permutation or time reversal, and generate non-trivial graphs different from threshold graphs and graphs produced by ${\bf R}_8$ (Fig.~\ref{ABgraphs}).  Finally, the rule ${\bf R}_{15}={1\,1\choose1\,1}$ is a degenerate
form of ${\bf R}_{13}$ (or ${\bf R}_{14}$) and yields only complete graphs (which are threshold graphs, so ${\bf R}_{15}$ is subsumed also in ${\bf R}_{4}$).

To summarize, ${\bf R}_4$, ${\bf R}_8$,  and ${\bf R}_{13}$ are the only two-letter rules that generate different classes of non-trivial connected graphs. 
There is yet another amusing type of symmetry: applying ${\bf R}_8$ and ${\bf R}_{13}$ to the same sequence yields {\it complement\/}, or {\it inverse\/ }graphs ---  nodes are adjacent in the inverse graph if and only if they are  {\it not\/} connected in the original graph.   The figure-background symmetry manifest in the rules 
${\bf R}_8$ and ${\bf R}_{13}$ ($0\leftrightarrow1$) is also manifest in the graphs they produce (Fig.~\ref{ABgraphs}a,c). 
On the other hand, the inverse of threshold graphs are also threshold graphs.  Also, the complement of
a threshold rule applied to the complement (inverted) sequence yields back the original graph. In this sense, threshold graphs 
have maximal symmetry.  ${\bf R}_8$-graphs are typically less dense, and ${\bf R}_{13}$-graphs are
typically denser than threshold graphs.

\begin{figure}[ht]
\includegraphics*[width=0.25\textwidth]{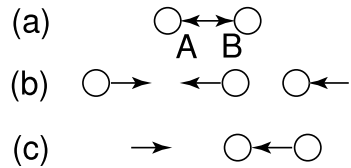}
\caption{Diagrammatic representation of rules for two-letter sequence nets:
(a)~All of the $2^2$ possible connections between nodes of type $A$ and $B$.
(b)~Three equivalent representations of the threshold rule ${\bf R}_4$.  The second and third diagram
are obtained by label permutation and time-reversal, respectively.
(c)~Diagrams for ${\bf R}_8$ and   ${\bf R}_{13}$. Note how they complement one another to the full
set of connections in part (a).}
\label{graph_notation}
\end{figure}

The connectivity rules have an additional useful interpretation as directed graphs, where the nodes
represent the letters of the sequence alphabet, a directed link, e,g., from $A$ to $B$ indicates the rule
$A\to B$, and a connection of a type to itself is denoted
by a self-loop (Fig.~\ref{graph_notation}).  Because the rules are the same under permutation of types, there is no need to actually
label the nodes: all graph isomorphs represent the same rule.  Likewise, time-reversal symmetry means that graphs with inverted arrows are equivalent as well. Note that the direction of self-loops is
irrelevant in this respect, so we simply take them as undirected.  We shall make use of this notation, extensively, for the analysis of 3-letter sequence nets in Section~\ref{three-letter}.  

\subsection{Alphabetical ordering}
%: example of eigenvalues of threshold nets (other?)
A very special property of sequence nets is the fact that any arbitrary ensemble of such nets possesses a natural ordering, simply listing the nets alphabetically according to their sequences.  In contrast, think for example of the ensemble of Erd\H os-R\'enyi random graphs of $N$ nodes, where links are present
with probability $p$: there is no natural way to order the $2^N$ graphs in the ensemble~\cite{ordering}.

\begin{figure}[ht]
\includegraphics*[width=0.45\textwidth]{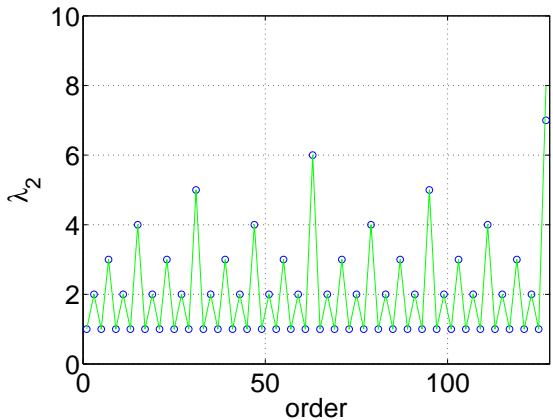}
\caption{Second smallest eigenvalues of threshold nets with $N=8$ nodes, plotted against their
alphabetical ordering.}
\label{eigs_2threshold}
\end{figure}

Plotting a structural property against the alphabetical ordering of the ensemble reveals some
inner structure of the ensemble itself, yielding new insights into the nature of the nets.  As an example,
in Fig.~\ref{eigs_2threshold} we show $\lambda_2$, the second smallest eigenvalue, for the
ensemble of connected threshold nets containing $N=8$ nodes (there are $2^7=128$ graphs in the ensemble, since their sequences must all start with the letter $A$).
Notice the beautiful pattern followed by the eigenvalues plotted in this way, which resembles 
a fractal, or a Cayley tree: the values within the first half of the graphs in the $x$-axis repeat in the second half, and the pattern iterates as we zoom further into the picture.

\subsection{The new classes of two-letter sequence nets}\label{new_classes}
%: Properties can be computed easily (some examples)
Structural properties of the new classes of two-letter sequence nets, ${\bf R}_8$ and ${\bf R}_{13}$, are as easily derived as for threshold nets.  Here we focus on ${\bf R}_8$ alone, which forms a subset of bipartite graphs.  The analysis for ${\bf R}_{13}$
is very similar and often can be trivially obtained from the complementary symmetry of the two classes.

All connected sequence nets in the ${\bf R}_8$ class must begin with the letter $A$ and end with
the letter $B$.  A sequence of this sort may be represented more compactly~\cite{hag} by the numbers of $A$'s and $B$'s in
the alternating layers, $(N_{A_1},N_{B_2},\dots,N_{B_n})$.  We assume that there are $N$ nodes and
$n$ layers ($n$ is even).  We also use the notation $N_A=\sum N_{A_i}$ and $N_B=\sum N_{B_i}$ for the total number of $A$'s
and $B$'s, as well as
\begin{equation}
N_{A_j}^-=\sum_{i<j}N_{A_i}\,;\qquad N_{A_j}^+=\sum_{i\geq j}N_{A_i}\,,
\end{equation}
and likewise for $N_{B_j}^\pm$.
Finally, since all the nodes in a layer have identical properties we denote any $A$ in the $i$-th layer
by $A_i$ and any $B$ in the $j$-th layer by $B_j$.  With this notation in mind we proceed to discuss
several structural properties.

\smallskip{\it Degree distribution\/}: Since $A$'s connect only to subsequent $B$'s (and $B$'s
only to preceding $A$'s) the degree $k$ of the nodes is given by
\begin{equation}
k(A_j)=N_{B_j}^+\,; \qquad k(B_j)=N_{A_{j}}^-\,.
\end{equation}

\smallskip{\it Clustering\/}:  There are no triangles in ${\bf R}_8$ nets so the clustering
of all nodes is zero.

\smallskip{\it Distance\/}:  Every $A$ is connected to the last $B$, so the distance between any two $A$'s is 2.  Every $B$ is connected to the first $A$ in the sequence, so the distance between any two
$B$'s is also 2.  The distance between  $B_i$ and $A_j$ is 1 if $j<i$ (they connect directly), and 3 if $j>i$
($B_i$ links to $A_1$, that links to $B_n$, that links to $A_j$).

\smallskip{\it Betweenness centrality\/}: Because of the time-reversal symmetry between $A$ and $B$, it suffices to analyze $B$ nodes only. The result for $A$ can then be obtained by simply reversing the creation sequence and permuting the letters.

The vertex betweenness $b(v)$ of a node $v$ is defined as:
\begin{equation}
	b(v) = \frac{1}{2}\sum_{s\neq t\neq v}{\frac{\sigma_{st}(v)}{\sigma_{st}}}
\end{equation}
where $\sigma_{st}$ is the number of shortest paths from node $s$ to $t$ ($s\neq t$), excluding the cases that $s=v$ or $t=v$. $\sigma_{st}(v)$ is the number of shortest paths from $s$ to $t$ that goes through $v$. The factor $\frac{1}{2}$ appears for undirected graphs since each pair is counted twice in the summation.

The betweenness of $B$'s can be calculated from lower layers to higher layers recursively.  In the first B-layer
\begin{equation}
	b(B_{2}) = \frac{\frac{1}{2}N_{A_1}(N_{A_1}-1)}{N_B}\,,
\end{equation}
and
\begin{equation}
\begin{split}
&	b(B_{j}) = b(B_{j-2}) \\
&\>\>\>\>\>
+ N_{A_{j-1}}\frac{\frac{1}{2}(N_{A_{j-1}}-1)+N_{A_{j-1}}^{-}}{N_{B_j}^+} + N_{A_{j-1}}\frac{N_{B_{j}}^-}{N_{B_j}^{+}}\,,
\end{split}
\end{equation}
for $j>2$.  The second term on the rhs accounts for the shortest paths from layer $A_{j-1}$ to itself and all previous layers of $A$, and the third term corresponds to paths from $A_{j-1}$ to $B_j$ to $A_i$ ($i<j-1$) to $B_{j-2}$.  Although this recursion can be solved explicitly it is best left in this form, as it thus highlights the fact
that the betweenness centrality increases from one layer to the next.  In other words, the networks are {\it modular\/}, where each additional $B$-layer dominates all the layers below.

\smallskip{\it Laplacian spectrum\/}: Unlike threshold nets, for ${\bf R}_8$ nets the eigenvalues
are {\it not\/} integer, and there seems to be no easy way to compute them.  Instead, we focus
on the second smallest and largest eigenvalues, $\lambda_2$ and $\lambda_N$, alone, for
their important dynamical role: the smaller the ratio $r\equiv\lambda_N/\lambda_2$ the more susceptible
the network is to synchronization~\cite{bar}.

\begin{figure}[ht]
\medskip
\includegraphics*[width=0.4\textwidth]{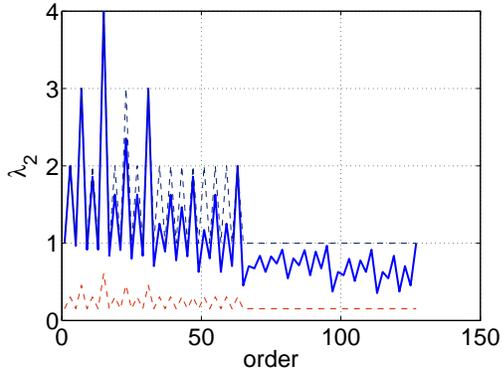}
\caption{Plot of second smallest eigenvalues of all connected $R_{8}$ nets with $N=8$ against their alphabetical ordering (solid curve), and their upper and lower bounds (broken lines).}
\label{l2bounds}
\end{figure}

Consider first $\lambda_2$.  For ${\bf R}_8$ it is easy to show that both the {\it vertex\/} and {\it edge connectivity\/} are equal to $\min(N_{A_1},N_{B_n})$.  Then, following an inequality in~\cite{moh},
\begin{equation}
2(1-\cos(\frac{\pi}{N}))\min(N_{A_1},N_{B_n})\leq\lambda_2\leq\min(N_{A_1},N_{B_n})\,.
\end{equation}
The upper bound seems stricter and is a reasonable approximation to $\lambda_2$ (see Fig.~\ref{l2bounds}).

\begin{figure}[ht]
\medskip
\includegraphics*[width=0.4\textwidth]{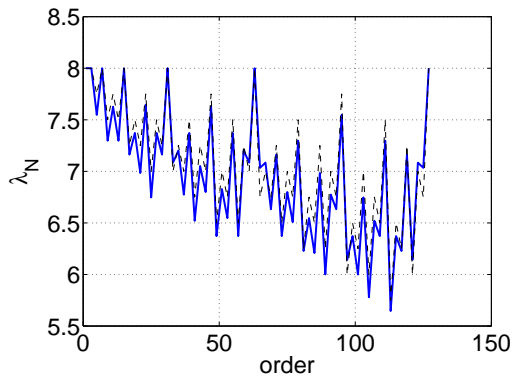}
\caption{Plot of largest eigenvalue of all connected $R_{8}$ nets with $N=8$ against their alphabetical ordering (solid curve), and its approximated value (broken line).}
\label{l2approx}
\end{figure}

For $\lambda_N$, using Theorem 2.2 of~\cite{moh} one can derive the bounds
\begin{equation}\
			\frac{N}{N-1}\max(N_{A},N_{B}) \leq \lambda_{N} \leq N\,,
\end{equation}
but they do not seems very useful, numerically.  Playing with various structural properties of
the nets, plotted against their alphabetical ordering, we have stumbled upon the approximation
\begin{equation}
	\lambda_{N}\mbox{ }\approx\mbox{ }N - \left(2\frac{N_{A}\cdot N_{B}}{N}-\left<k\right>\right)\,,
\end{equation}
where $\left<{k}\right>$ is the average degree of the graph, see Fig.~\ref{l2approx}. The approximation is exact for bipartite {\it complete\/}
graphs ($n=1$) and the relative error increases slowly with $N$; it is roughly at 10\% for $N=60$.

\subsection{Relation to threshold nets}
%: Relation to threshold nets
In~\cite{hag} it was shown that threshold graphs have a mapping to a sequence net, with a unique sequence (under the ``threshold rule" ${\bf R}_4$); and conversely,  for any ${\bf R}_4$-sequence net 
there exists a set of weights  $x_i$ of the nodes (not necessarily unique), such that connecting any two nodes that satisfy $x_i+x_j>\theta$ reproduces the sequence net.   Here we establish a similar relation
between ${\bf R}_8$- (or ${\bf R}_{13}$-) sequence nets and a different kind of threshold net, where
connectivity is decided by the difference $|x_i-x_j|$ rather than the sum of the weights.

We begin with the mapping of a weighted set of nodes to a ${\bf R}_8$-sequence net.  Let
a set of $N$ nodes have weights $x_i$ ($i=1,2,\dots,N$), taken from some probability density, and we assume $0< x_i<2\theta$, without
loss of generality.  Denote nodes with $x_i<\theta$ as type $A$ and nodes with $x_i>\theta$ as type $B$.
Finally, connect any two nodes $i$ and $j$ that satisfy $|x_i-x_j|>\theta$.  The resulting graph can be
constructed by a unique sequence under the rule ${\bf R}_8$, obtained as follows.

For convenience, rewrite the set of weights as
\begin{equation}
0<u_1<u_2\cdots< u_{N_A}<\theta<v_1<\cdots<v_{N_B}<2\theta\,,
\end{equation}
where the first $N_A$ weights correspond to $A$-nodes and the rest to $B$-nodes.
Denote the creation sequence by $(S_1,S_2,\dots,S_N)$ and determine the $S_i$ by the algorithm
(in pseudo-code):

\medskip\noindent
{\tt Set} $i=1$, $j=1$

\noindent
{\tt For} $k=1,2,\dots,N$, {\tt do:}

\noindent\hskip 0.4cm
{\tt If} $|u_i-v_j|>\theta$
         
\noindent\hskip 0.8cm         
{\tt set} $S_k=A$ {\tt and} $i=i+1;$

\noindent\hskip 0.4cm
{\tt Else}

\noindent\hskip 0.8cm         
{\tt set} $S_k=B$ {\tt and} $j=j+1.$

\noindent
{\tt End.}

\medskip\noindent
It is understood that if the $u_i$ are exhausted before the end of the loop, the remainder $B$-nodes are
automatically affixed to the end of the sequence (and similarly for the $v_j$).
For example, using this algorithm we find that the ``difference-thre\-shold" graph resulting from the set of weights
$\{$1,2,3,5,7,16,17,20$\}$ and $\theta=12$, can be reproduced from the sequence
$(A,A,A,B,B,A,A,B)$, with the rule ${\bf R}_8$.

Consider now the converse problem: given a graph created from the sequence $(S_1,S_2,\dots,S_N)$ with the rule
${\bf R}_8$, we derive a (non-unique) set of weights $\{x_i\}$ such that connecting any two nodes with
$|x_i-x_j|>\theta$ results in the same graph.  Rewrite first the creation sequence into its compact form
$(N_{A_{1}},N_{B_{2}},...,N_{B_{n}})$,
and assign weights $l$ for nodes $A$ in layer $l$, weights $n+m$ for nodes $B$ in layer $m$, and
set the threshold at $\theta=n$. For example, the sequence $(A,A,A,B,B,A,A,B)$ has a compact representation $(3,2,2,1)$, with $n=4$ layers, so the three $A$'s in layer $1$ have weights $1$, the two $B$'s in layer $2$ have weights $6$, the two $A$'s in layer $3$ have weights $3$, and the single $B$ in layer $4$ has weight $8$. The weights $\{1,1,1,6,6,3,3,8\}$, with connection threshold $\theta = 4$, reproduce the original graph.

Sequence graphs obtained from the rule ${\bf R}_{13}$ can be also mapped to difference threshold graphs in
exactly the same way, only that the criterion for connecting two nodes is then $|x_i-x_j|<\theta$, instead of
$|x_i-x_j|>\theta$, as for ${\bf R}_8$.  The mapping of sequence nets to generalized threshold graphs may
be helpful in the analysis of some of their properties, for example, for finding the {\it isoperimetric number\/}
of a sequence graph~\cite{moh,isoperimetric}. 

\section{Three-Letter Sequence Nets}
\label{three-letter}         
%: Classification of 3-letter sequence nets
\subsection{Classification}

With a three-letter alphabet, $\{A,B,C\}$, there are at the outset $2^{3^2}=512$ possible rules.
Again, these can be reduced considerably, due to symmetry.  Because the rule matrix has 9 entries
(an odd number) no rule can be identical to its complement.  Thus, we can limit ourselves
to rules with no more than 4 non-zero entries and apply symmetry arguments to reduce their space
--- at the very end we can then add the complements of the remaining rules.

\begin{figure}[ht]
\medskip
\includegraphics*[width=0.4\textwidth]{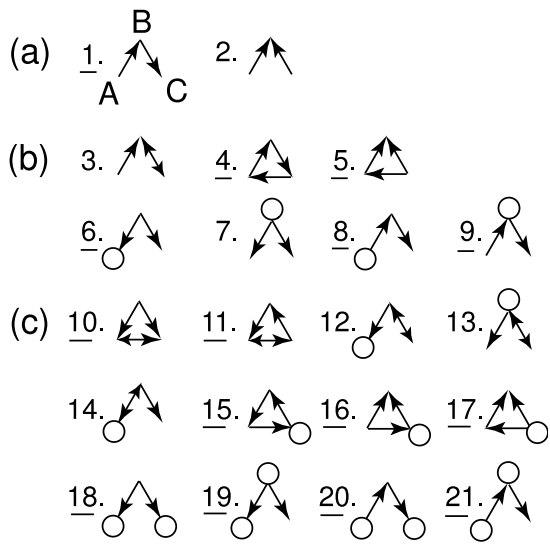}
\caption{Rules for three-letter sequence nets:
Shown are rules with (a)~two, (b)~three, and (c)~four interactions.  All label permutations and time reversals
are omitted.  In addition, rules 2 and 7 degenerate to two-letter rules (identifying $A$ and $C$), and rules 3, 12, 13, and 14 are degenerate
cases of rules 2, 6, 7, and 6, respectively.  This leaves us with fifteen distinct three-letter rules (underlined), and their fifteen complements, for a total of 30 different classes of three-letter sequence nets.}
\label{3nets}
\end{figure}

In Fig.~\ref{3nets} we list all possible three-letter rules with two, three, and four interactions.  Rules that lead to disconnected graphs, and symmetric rules (by label permutation or time-reversal) have
been omitted from the figure.  

Rule \Re2~\cite{rem3} is in fact not new: identifying nodes of type $A$ and $C$ (as marked
in rule 1 of the figure) we can easily see that the rule is identical to the two-letter rule \Ro8.  In the same fashion,
rule \Re7 is the same as the two-letter threshold rule \Ro4.  

Rule \Re3 is a degenerate form of  \Re2: Because of the double connection $B\to C$ and $C\to B$, the order at which $B$ and $C$ appear in the sequence relative to one another is inconsequential.  (On the other hand, the order of the $B$'s relative to $A$'s {\it is\/} important, since $A$'s connect only to those $B$'s that appear earlier in the sequence.)  Then, given a sequence one can rearrange it by moving all the $C$'s to the end of the list.
If we now apply \Re2, $A\to B$ and $C\to B$, then we get the same graph as from the original sequence under the rule \Re3.  The same consideration applies to rules \Re{12}, \Re{13} and \Re{14}, that are degenerate forms of \Re6, \Re7 and \Re8 (or \Re6), respectively.  We are thus left with only 15 distinct rules with fewer than 5 connections.  To
these one should add their complements, for a total of 30 distinct three-letter rules.  

Note the resemblance of \Re{9}, \Re{18}, and \Re{20} to two-letter threshold nets.  \Re{18} seems like a particularly symmetrical generalization and we will focus on it in much of our discussion below.

\subsection{Connectedness}\label{connect}
While one can easily establish
wether a graph is connected or not, {\it a posteriori\/}, with  a burning algorithm that requires ${\cal O}(N)$ steps,
it is useful to have shortcut rules that tell us how to avoid bad sequences at the outset: knowing that two-letter threshold graphs are connected if and only if their sequence ends with
$B$, deals with the question most effectively.  Analogous criteria exist for three-letter sequence graphs
but they are a bit more complicated.
For example, three-letter sequences interpreted with \Re{18} lead to connected graphs if and only if they satisfy:  {\it (1)~The first {\rm A} and the first {\rm C} in the sequence appear before the last {\rm B}. (2)~The sequence does not start with {\rm B}}.   (We assume that the sequence contains all three letters.) For \Re1the requirements are:
{\it
(1)~The first {\rm A} in the sequence must appear after the first {\rm B}.
(2)~The last {\rm C} in the sequence must appear before the last {\rm B}.
(3)~The last {\rm A} in the sequence must appear after the first {\rm C}, and there ought to be at least one {\rm B} between the two.}  Similar criteria exist for all other three-letter rules and can be found by inspection.

\subsection{Structural properties}
Structural properties of three-letter sequence nets are analyzed as easily as those of two-letter nets,  Here we
list, as an example, a few basic attributes of \Re{18} sequence nets.  We use a notation similar to that of Section~\ref{new_classes}.

\smallskip{\it Degree distribution\/}:
$A$ and $C$ nodes form complete subgraphs, while $B$ nodes connect to all preceding $A$'s and $C$'s. Thus the degree of the nodes are:
\begin{equation}
\begin{split}
	&k(A_i) = N_{A} - 1 + N_{B_i}^{+}\,, \\
	&k(B_i) = N_{A_i}^{-} + N_{C_i}^{-}\,, \\
	&k(C_i) = N_{C} - 1 + N_{B_i}^{+}\,.
	\end{split}
\end{equation}

\begin{figure}[ht]
\medskip
\includegraphics*[width=0.4\textwidth]{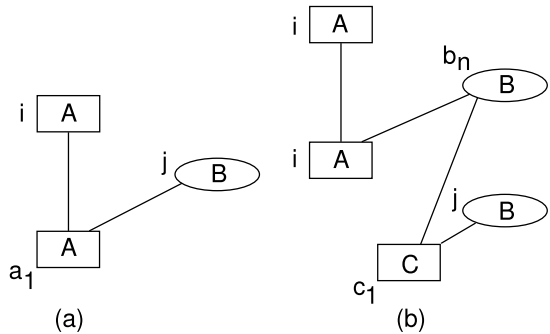}
\caption{The distance $d(A_i,B_j)$ in \Re{18} nets. (a)~If $i>j$ and the first $A$ is below $B_j$ the distance is 2. (b)~If the first $A$ is above $B_j$, then the first $C$ must be below ($B$ can't start the sequence); in that case if $A_i$ is below the last $B$ the distance is 3, and otherwise the distance is 4. Only the relevant parts of the complete net are shown.}
\label{distance}
\end{figure}

\smallskip{\it Distance\/}:
Since the $A$ nodes make a subset complete graph $d(A_i,A_j)=1$, and likewise for $C$, $d(C_i,C_j)=1$.
The $B$'s do not connect among themselves, but they all connect to the nodes in the first layer (which does
not consist of $B$'s), so $d(B_i,B_j)=2$.
For the distance of $A$ nodes from $B$, we have
\begin{equation}
	  d(A_i,B_j) = 
	  \begin{cases}
	  1 & i<j\,,\\
	  2  & i>j, \,a_1<j\,, \\
	  3 & i>j, \,a_1>j,\, i<b_n\,,\\
	  4 & i>j,\, a_1>j,\, i>b_n\,,
	  \end{cases}
\end{equation}	  
where $a_1$ is the index of the first $A$-layer and $b_n$ is the index of the last $B$-layer.
The first line follows since $B$'s are directly connected to preceding $A$'s and $C$'s. The second, and third and fourth lines are illustrated in Fig.~\ref{distance}a and b, respectively.  The distance $d(C_i,B_j)$ follows the very same pattern.  Finally, inspecting all different cases one finds
\begin{equation}
	  d(A_i,C_j) = 
	  \begin{cases}
	  2 & i,j<b_n\,,\\
	  3  & i<b_n<j, {\rm\ or\ }j<b_n<i\,, \\
	  4 & i,j>b_n\,.	  \end{cases}
\end{equation}	

\begin{figure}[ht]
\medskip
\includegraphics*[width=0.4\textwidth]{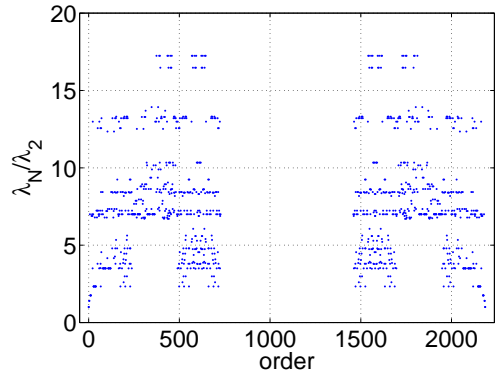}
\caption{The ratio $\lambda_N/\lambda_2$ for \Re{18} nets consisting of $N=7$ nodes, against their alphabetical ordering.  Note
the gap near the center, which corresponds to sequences of disconnected graphs.  Note also the mirror symmetry --- this is due to the mirror symmetry of the rule \Re{18} itself.}
\label{r_R18}
\end{figure}

\smallskip{\it Eigenvalues\/}:
We have found no obvious way to compute the eigenvalues, despite the similarities between 
\Re{18} nets and two-letter threshold nets.  However, plots of the eigenvalues against the alphabetical
ordering of the nets once again reveals intriguing fractal patterns, and one can hope that these might be exploited at the very least to produce good bounds and approximations.  In Fig.~\ref{r_R18} we plot
the ratio $r=\lambda_N/\lambda_2$ for \Re{18} nets with $N=7$ against their alphabetical ordering.
The $x$-axis includes sequences of nets that are not connected: In this case $\lambda_2=0$ and synchronization is not possible.  These cases show as gaps in the plot, for example, the big gap in the center
corresponds to disconnected sequences that start with the letter $B$ (see Section~\ref{connect}).

\subsection{Multi-threshold nets}
Some of the three-letter sequence nets can be mapped to generalized forms of threshold nets.
For example, the following scheme yields a {\it two\/}-threshold net, equivalent to three-letter sequence nets
generated by the rule \Re{20}.  Let the nodes be assigned weights $0<x_i<3\theta/2$, from a random distribution, and connect any two nodes $i$ and $j$ that satisfy $x_i+x_j<\theta\equiv\theta_1$ or
$x_i+x_j>2\theta\equiv\theta_2$.  Identifying nodes with weight $0<x_i<\theta/2$ with $A$, nodes with
$\theta/2<x_i<\theta$ with $B$, and nodes with $\theta<x_i<3\theta/2$ with $C$, we see that all $A$'s
connect to one another and all $C$'s connect to one another but the $B$'s do not, and $A$'s and $C$'s do not connect;  nodes of type $A$ and $B$ may or may not connect, and likewise for nodes of type $C$ and $B$.
To reflect the actual connections, the nodes of type $A$ and $B$ may be arranged in a sequence according
to the algorithm in~\cite{hag}, for the threshold rule \Ro{5}.  Also the nodes of type $C$ and $B$ may be
arranged in a sequence, to reflect the actual connections, with the very same algorithm.  Because there
are no connections between $A$ and $C$ the two results may be trivially merged.  Note, however, that
once the $A$-$B$ sequence is established the order of the $B$'s is set, so the direction of connections between $C$ and $B$ ($C\to A$ or
$A\to C$) is {\it not\/} arbitrary.  In our example, the mapping is possible to \Re{20} but not to \Re{18}.

\section{summary and discussion}
\label{conclude}

We have introduced a new class of nets, sequence nets, obtained from a sequence of letters and fixed rules of connectivity.  Two-letter sequence nets contain threshold nets, and in addition two newly discovered classes.
The \Ro{13} class can be mapped to a ``difference-threshold" net, where nodes $i$ and $j$ are connected
if their weights difference satisfies $|x_i-x_j|<\theta$.  This type of net may be a particularly good model for
social nets, where the weights might measure political leaning, economical status, number of offspring, etc., and
agents tend to associate when they are closer in these measures.  We have shown that the structural properties of
the new classes of two-letter sequence nets can be analyzed with ease, and we have introduced an ordering in ensembles of sequence nets
that is useful in visualizing and studying their various attributes.

We have fully classified 3-letter sequence nets, and looked at a few examples, showing that they too can be analyzed simply.  
The diameter of sequence nets grows linearly with the  number of letters in the alphabet and for a 3-letter alphabet
it is already 3 or 4, comparable to many everyday life complex nets.   Realistic diameters might be achieved with a modest expansion of the alphabet. 

There remain numerous open questions: Applying symmetry arguments we have managed to reduce the class of 3-leter nets to just 30 types, but we have not ruled out the possibility that some overlooked symmetry might reduce the list further;  The question of which sequences lead to connected nets can be studied by
inspection for small alphabets, but we have no comprehensive approach to solve the problem in general; We have shown how to map sequence nets to generalized types of threshold nets, in some cases ---  Is such a mapping always possible?  Is there a systematic way to find such mappings for any sequence rule?;  What kinds of nets would
result if the connectivity rules applied only to the $q$ preceding letters, instead of to {\it all\/} preceding letters? etc.
We hope to tackle some of these questions in future work.

\acknowledgments Partial funding from the NSF (DbA) and ARO (JS) is gratefully acknowledged.

%:REFERENCES
%


\begin{thebibliography}{99}
\bibitem{boxy} We prefer the more ``boxy" look of rectangular boxes to denote the tight connections
in complete graph cliques.
\bibitem{rem2}
Here ${\bar S}_n$ stands for the inverted type: ${\bar S}_n=A$ if $S_n=B$, and vice versa.
\bibitem{ordering} One could order the graphs partially, according to their size (the number of links),
however, it is less clear how to order  the  $\left({N\atop m}\right)$ ``degenerate" graphs with $m$ links.
\bibitem{isoperimetric} The isoperimetric number of a graph $G$ is $i(G)=\inf_X|\partial X|/|X|$,
where the infimum is taken over all subsets $X$ of $V(G)$ satisfying $|X|\leq\frac{1}{2}|V(G)|$.
\bibitem{rem3} We use the superscript to distinguish between two- and three-letter rules when doubt
may arise otherwise.
\end{thebibliography}
\end{document}